# Grating-coupling-based excitation of Bloch surface waves for lab-on-fiber optrodes

**MICHELE SCARAVILLI,**[1,2] **GIUSEPPE CASTALDI,**[1] **ANDREA CUSANO,**[2,3] **AND VINCENZO GALDI**[1,4]

[1]*Waves Group, Department of Engineering, University of Sannio, I-82100, Benevento, Italy*
[2]*Optoelectronic Division, Department of Engineering, University of Sannio, I-82100, Benevento, Italy*
[3]*a.cusano@unisannio.it*
[4]*vgaldi@unisannio.it*

**Abstract:** In this paper, we investigate the possibility to excite Bloch surface waves (BSWs) on the tip of single-mode optical fibers. Within this framework, after exploring an idealized, proof-of-principle grating-coupling-based scheme for on-tip excitation of BSWs, we focus on an alternative configuration that is more robust with respect to fabrication-related non-idealities. Subsequently, with a view towards label-free chemical and biological sensing, we present a specific design aimed at enhancing the sensitivity (in terms of wavelength shift) of the arising resonance with respect to changes in the refractive properties of the surrounding environment. Numerical results indicate that the attained sensitivities are in line with those exhibited by state-of-the-art plasmonic bioprobes, with the key advantage of exhibiting much narrower spectral resonances. This prototype study paves the way for a new class of miniaturized high-performance surface-wave fiber-optic devices for high-resolution label-free optical biosensing, and represents an important step forward in the "lab-on-fiber" technology roadmap.

## References and Links

## 1. Introduction

The possibility to perform real-time, label-free selective detection of several chemical and biological species by means of specifically designed photonic devices makes optical biosensing of paramount interest for potential applications in diverse areas, including biomedical research, healthcare, homeland security and environmental monitoring [1,2].

To perform high-sensitivity label-free biomolecular analyses, a resonant structure supporting electromagnetic surface waves (SWs) represents an intrinsically ideal tool, as it allows light confinement at nanoscale and thus the exploitation of a strong light-matter interaction in close proximity of the interface between a properly designed sensitive element and the external medium, i.e., the fluid containing the target molecules to be detected [3].

An electromagnetic SW can be seen as a guided mode travelling in a direction parallel to the interface between two media with suitably chosen optical properties [3]. Depending on the material properties, different SWs with specific characteristics can be obtained, such as surface plasmon polaritons (SPPs) [3,4], sustained by metallo-dielectric interfaces, and Dyakonov waves [3,5], which can propagate at the interface between an isotropic and an anisotropic dielectric material.

SPP excitation is the physical principle behind the operation of surface plasmon resonance (SPR) biosensors, which represent the standard label-free SW sensors [6,7]. On the other hand, the optical losses associated to common plasmonic materials (e.g., noble metals) yield a characteristic *broadening* of the arising spectral resonances, with severe impact on the final limit of detection (LOD) [8]. As the field of plasmonics has reached the "slope of enlightenment" [9,10], consensus is generally growing on the quest for alternative low-loss materials and related mechanisms.

For instance, Dyakonov-wave-based sensors, which could in principle rely solely on transparent materials, may provide an attractive alternative. Unfortunately, such waves can be excited and observed under very stringent conditions. For this reason, although they were theoretically predicted and studied few decades ago, their first experimental demonstration was only recently reported [11], and different design strategies [12] and configurations [13] have been explored to improve the visibility of the phenomenon.

A more practically viable route is constituted by Bloch SWs (BSWs), which have recently elicited renewed attention in connection with label-free biosensing [14–24], even though they have been known theoretically [25] (and verified experimentally [26]) since the 1970s. A BSW is guided at the interface between a homogeneous medium and a truncated one-dimensional photonic crystal (1DPC), and is characterized by a dispersion curve located in the forbidden band of the 1DPC [3,25]. By comparison with the plasmonic counterpart, a 1DPC exhibits a reduced confinement capability of the SW evanescent tail, yielding a lower rate of SW power in the external medium, which corresponds to lower values of the sensitivity to changes of its refractive properties [18,19]. Nonetheless, the reliance on low-loss materials (similar as the Dyakonov-wave case above) implies *much narrower* resonances, which lead to larger *overall* figures of merit (FOMs) [17–21]. Moreover, the photonic-crystal-based technological platform offers other advantages in terms of increased design flexibility and wavelength scalability [18–20,27].

This has led to the development of BSW-based label-free biosensors exhibiting promising performance (in terms of LOD) in a wide range of applications, including antigen-antibody interactions [16,22–24], DNA binding [16] and protein aggregation [15], just to mention a few.

The present study represents a first step towards the integration of the BSW-based sensing mechanism in the emerging paradigm of "lab-on-fiber" technology, which entails novel classes of "all-fiber" multi-functional devices arising from the integration of highly functional micro- and/or nano-structured materials directly on the tip of optical fibers [28–30]. This brings about a series of inherent advantages, such as reduced footprint and weight, compactness, flexibility, mechanical robustness, easy integration in hypothermic needles for *in vivo* analyses, and easy integration of complex illumination systems and sophisticate detection modules without any need to preserve careful alignment [28–30].

Within this framework, in spite of several demonstrations in planar structures [14–24], the on-fiber integration of BSW-type structures has received very little attention. In fact, while the integration of 1DPCs on the tip of an optical fiber does not pose particularly severe technological challenges, the excitation mechanism represents a pivotal issue. Indeed, BSW-based devices typically rely on the well-known Kretschmann-Raether prism-coupled configuration [14–24] in order to realize the appropriate phase-matching conditions. However, this type of configuration is inherently unsuited to lab-on-fiber optrodes, and alternative mechanisms are required to select the correct incident conditions for BSW excitation.

To this aim, a fiber-optic label-free evanescent-wave scheme based on a 1DPC-coated D-type fiber was theoretically proposed in [31]. However, a practical implementation would pose challenges in the control of the modal distribution within the active region of the device, as well as the polarization cross sensitivity typical of fiber-optic structures that are not ring-shaped symmetric.

In this work, we propose an alternative excitation scheme based on a grating-coupling mechanism, which enables a more efficient control of the excitation angle combined with a feasible integration on the tip of an optical fiber.

Accordingly, the rest of the paper is organized as follows. In Sec. 2, we present a first configuration suitable to simply demonstrate the possibility of coupling a BSW on the tip of a single-mode optical fiber via a 1D diffraction grating, but revealing an intrinsic weakness against predictable fabrications tolerances. In Sec. 3, we overcome this problem by proposing an alternative scheme and exploring in detail the several degrees of freedom available. In Sec. 4, we introduce a design aimed at improving the behavior of the structure as a refractive-index sensor, and compare the attainable sensitivity with those exhibited by state-of-the-art plasmonic fiber-tip bioprobes. In Sec 5, we evaluate the robustness of the presented results against material losses in the 1DPC. Finally, in Sec. 6, we provide some brief concluding remarks and hints for future research.

## 2. On-tip diffraction-grating configuration

*2.1 Geometry and generalities*

The most critical issue to face in our design is the non-zero angle of incidence required for the excitation of a SW at the interface between a 1DPC deposited on the fiber tip and an external medium [3]. In a single-mode optical fiber, the fundamental mode propagates along the core axis, leading to normal incidence at the fiber tip. While the required oblique incidence might be, in principle, attained by considering *higher-order* modes (i.e., a multimode fiber), controlling the multimode interactions and coupling with the SW would be very complicated.

To overcome the normal incidence issue in a single-mode fiber, we first analyze the idealized grating-coupled configuration schematized in Fig. 1. In particular, we assume an operational wavelength $\lambda_0 = 1550\,\text{nm}$, and consider silicon monoxide (SiO, with refractive index $n_1 = 1.8773$) and silica (SiO$_2$, $n_2 = 1.4657$) as high- and low-contrast materials, respectively, for the 1DPC and for the grating [32]. The 1DPC is terminated with a SiO layer, and the external medium is air ($n_{ext} = 1$). As customary in lab-on-fiber scenarios [28,29,33–36], in order to guarantee the computational affordability of our numerical study, we approximate the impinging fiber mode with a normally-incident plane wave, thereby neglecting the illumination tapering and polarization non-uniformity. It is worth emphasizing that previous experimental validations in other lab-on-fiber configurations have indicated that this approximation typically provides acceptable accuracy in the prediction of the optical response and sensitivity [28,33–36].

Via a judicious design of the coupling grating, an efficient power transfer between the incident light and scattered beams at specific design angles $\pm\bar{\theta}_s$ is possible.

Grating-coupling-based excitation of BSWs was theoretically studied in [3,37], with focus on the excitation of multiple BSWs via an angular interrogation scheme. Coupling at normal incidence was initially considered in [38], to demonstrate the diffraction-assisted BSW-coupled fluorescence extraction principle by means of linear and circular gratings. In [39], subwavelength focusing of BSWs on a circular grating starting from normal incidence was investigated. Applications related to label-free biosensing can be found in [40,41]. In [40], an angular interrogation of a porous silicon 1DPC for the simultaneous excitation of a BSW and a sub-surface confined optical mode was explored. In [41], a numerical analysis of a wavelength-interrogated configuration similar that in Fig. 1 was carried out. However, all the above results are not directly relevant for our purpose of a feasible and efficient on-fiber integration of a BSW-based sensing device, and a specific study is therefore needed.

*2.2 BSW excitation*

For the generation of an evanescent wave in the external medium, the coupling angle $\bar{\theta}_s$ needs to exceed the silica-air critical angle $\theta_c = \arcsin(n_{ext}/n_{inc}) \approx 43.02^\circ$. As a mere proof-of-principle of the scheme in Fig. 1, we consider $\bar{\theta}_s = 60^\circ$. Accordingly, the 1DPC can be dimensioned as a quarter-wavelength stack [27] for an incidence angle $\bar{\theta}_s$ at the operational wavelength ($\lambda_0 = 1550\,\text{nm}$). The layer thicknesses $d_1$ and $d_2$ are thus given by

$$d_1 = \frac{\lambda_0}{4n_1 \cos\theta_1} \approx 280\,\text{nm}, \tag{1}$$

$$d_2 = \frac{\lambda_0}{4n_2 \cos\theta_2} \approx 529\,\text{nm}. \tag{2}$$

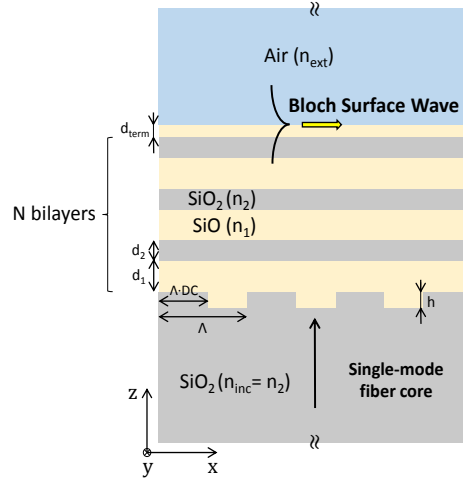

Fig. 1. 2D schematic of the on-tip diffraction grating idealized configuration in the ($x$-$z$) reference system, with geometry and field quantities assumed as invariant along the $y$-direction. A 1DPC made of $N$ SiO/SiO$_2$ bi-layers (with thicknesses $d_1$ and $d_2$, respectively) is deposited on the tip of a single-mode optical fiber, and terminated with a SiO layer of thickness $d_{term}$. The coupling is mediated by a SiO/SiO$_2$ diffraction grating with period $\Lambda$, height $h$, and duty-cycle $DC$, placed on the fiber tip (at the bottom of the 1DPC). The external medium is assumed as air.

Finally, the grating period $\Lambda$ readily follows from the phase matching condition, viz.,

$$m\frac{2\pi}{\Lambda} = \bar{k}_x. \tag{3}$$

with $m$ denoting an integer number. Choosing $m=1$, i.e., selecting the first Bragg diffraction orders as the excitation beams (propagating along the $\pm\bar{\theta}_s$ directions), yields $\Lambda \approx 1221\,\text{nm}$.

The remaining grating geometric parameters (height $h$ and duty-cycle $DC$) as well as the number of 1DPC bilayers ($N$) do not affect the coupling wavelength, and can be selected so as to optimize the characteristics of the arising resonance in terms of visibility in the spectral range of analysis and linewidth.

### 2.3 Optical response and effects of the fabrication-related non-idealities

In Fig. 2(a), as proof of principle, we show a zeroth-order reflectance spectrum, numerically computed via full-wave simulations based on the rigorous coupled-wave approach (RCWA) [43,44] for a TE-polarized incident plane-wave, and corresponding to light travelling back along the fiber axis after reflection from a structure with $h = 300\,\text{nm}$, $DC = 0.5$ and $N = 5$. Around the operational wavelength $\lambda_0 = 1550\,\text{nm}$, it exhibits the expected BSW resonance superposed to a Fabry-Perot-type background; this latter arises from the interaction between the zeroth-order mode and the 1DPC layers. The resonance exhibits a characteristic asymmetric Fano-like lineshape, typically encountered in scattering problems featuring the interaction between continuous and discrete modal spectra [45]. This is not surprising, recalling that, in our configuration, the coupling of the incident light (radiation continuum) with the BSW is mediated by the discrete Bragg spectrum of the periodic diffraction grating. We estimated, via a standard Breit-Wigner-Fano (BWF) numerical fit [46], a linewidth $\Gamma$ of $\sim 0.37\,\text{nm}$ and a $Q$-factor of $\sim 4200$. As clearly observable in the field map shown in Fig.

2(b), the resonance is accompanied by strong field localization at the interface between the 1DPC and air.

Although the proposed structure is able, at least in principle, to provide a suitable method for BSW excitation in the case of normal incidence, unavoidable effects related to the fabrication process need to be taken into account. Even though the diffraction grating can be efficiently written on the tip of an optical fiber by using a focused ion beam (FIB) milling procedure recently demonstrated in [33], its corrugated shape will partially "propagate", thereby perturbing the subsequently deposited 1DPC multilayer. A realistic quantitative model of the relief-propagation phenomenon is difficult to gauge *a priori*. Nevertheless, to roughly estimate the influence of such effect, we numerically investigated a structure in which all the 1DPC layers exhibited a periodic undulation characterized by same period and duty cycle of the on-tip diffraction grating, and height of 150 nm (i.e., 50% of the original grating height). As it can be observed in Figs. 2(c) and 2(d), although the BSW excitation still occurs, there is a sensible bandwidth increase up to one order of magnitude in $\Gamma$ ($\sim 8.7\,\text{nm}$). This leads to a significant detrimental effect on the FOM of the final device, with a decrease of the $Q$-factor down to $\sim 180$.

To overcome this important issue, while maintaining the fundamental advantage of narrowband excitation, in what follows, we propose and explore and alternative configuration.

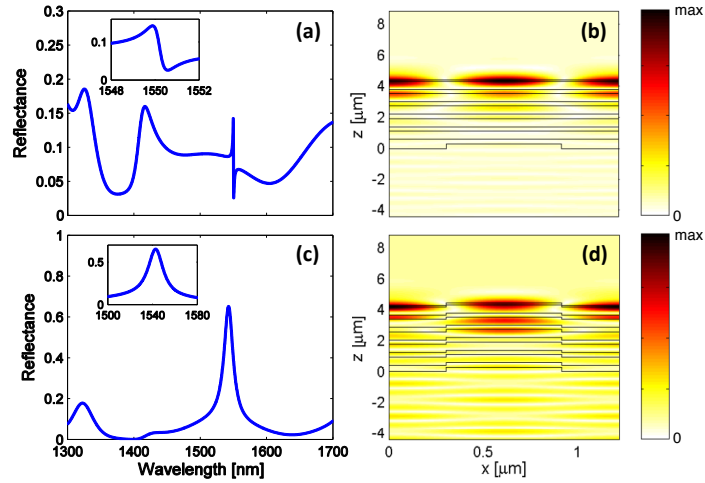

Fig. 2. (a) Zeroth-order reflectance spectrum for a structure with $d_1 = 280\,\text{nm}$, $d_2 = 529\,\text{nm}$, $d_{term} = 92\,\text{nm}$, $N = 5$, $\Lambda = 1221\,\text{nm}$, $h = 300\,\text{nm}$ and $DC = 0.5$. In the inset, the region around the BSW resonance is magnified. (b) Electric field magnitude map within a unit cell, computed at the operational wavelength $\lambda_0 = 1550\,\text{nm}$. (c) Same as panel (a), but taking into account relief-propagation effects: all 1DPC layers are assumed to exhibit a periodic square wave undulation characterized by period $\Lambda$, duty cycle $DC$ and height $h/2$. (d) As in panel (b), but for the structure pertaining to panel (c).

## 3. Alternative diffraction-grating configuration

Referring to the schematic in Fig. 3(a), the 1D diffraction grating is now placed on the surface of the 1DPC, which is directly deposited on the fiber tip, thereby avoiding any detrimental effect related to relief propagation through the 1DPC structure. Also in this case, the operational wavelength $\lambda_0$ and the pair of materials for the 1DPC are assumed as in Sec. 2.

## 3.1 BSW excitation

By assuming also in this case $\bar{\theta}_s = 60°$, Eqs. (1)-(3) yield once again: $d_1 = 280\,\text{nm}$, $d_2 = 529\,\text{nm}$ and $\Lambda = 1221\,\text{nm}$. The undulated termination layer parameters ($d_t$, $h$ and $DC$) need to be dimensioned in order for the structure to sustain a BSW at the operational wavelength as well as for the tuning of the arising BSW resonance. As before, $N$ represents a further degree of freedom for the optimization of the resonant lineshape, and it does not affect the coupling wavelength.

In Fig. 3(b), we show the zeroth-order reflectance spectrum for a structure with $d_t = 50\,\text{nm}$, $h = 100\,\text{nm}$, $DC = 0.5$ and $N = 5$. A highly visible and rather broad peak ($\Gamma \approx 6\,\text{nm}$) arises around $\lambda_0$. The field-map in Fig. 3(c), evaluated at $\lambda_0$, closely resembles that in Fig. 2(b), and proves the correct excitation of a BSW at the undulated interface between the 1DPC and the external medium.

We point out that the average thickness of the assigned termination layer is $100\,\text{nm}$, close to the value calculated for $d_{term}$ in the previous configuration ($92\,\text{nm}$). Moreover, we verified in different cases that, for a weak surface corrugation ($h \ll \lambda_0$), the solution of the dispersion equation given in [42] provides a fairly good estimate of the average thickness of the termination layer necessary for sustaining the BSW, given the design parameters $(\lambda_0, \bar{\theta}_s)$.

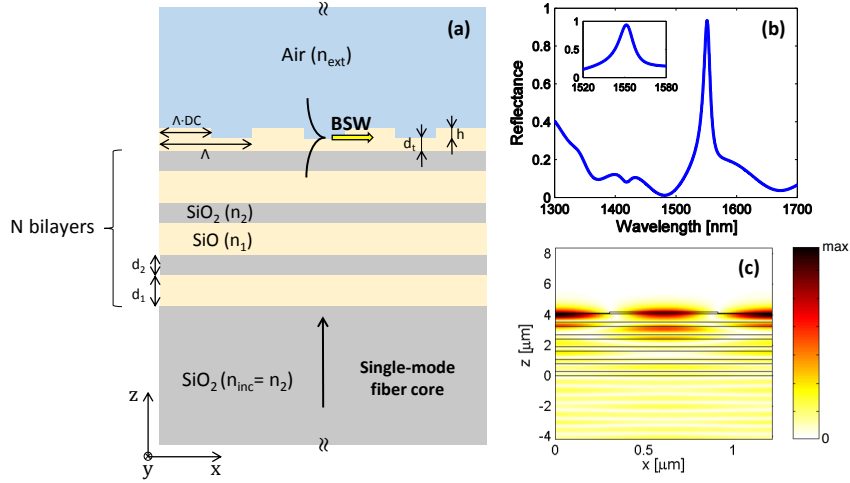

Fig. 3. (a) 2D schematic of the on-tip diffraction-grating alternative configuration in the ($x$ - $z$) reference system, with geometry and field quantities assumed as invariant along the $y$-direction. A 1DPC made of $N$ SiO/SiO$_2$ bi-layers (with thicknesses $d_1$ and $d_2$, respectively) is deposited on the tip of a single-mode optical fiber, and terminated with a SiO layer of thickness $d_t$. The coupling is mediated by a SiO/air diffraction grating with period $\Lambda$, height $h$, and duty-cycle $DC$, placed on the top of the 1DPC. The external medium is assumed as air. (b) Zeroth-order reflectance spectrum for a structure with $d_1 = 280\,\text{nm}$, $d_2 = 529\,\text{nm}$, $d_t = 50\,\text{nm}$, $N = 5$, $\Lambda = 1221\,\text{nm}$, $h = 100\,\text{nm}$ and $DC = 0.5$. In the inset, the region around the BSW resonance is magnified. (c) Electric field magnitude map within a unit cell, computed at the operational wavelength $\lambda_0 = 1550\,\text{nm}$.

### 3.2 Tailoring of the BSW resonance lineshape

As previously mentioned, the BSW resonant lineshape can be properly tailored by exploiting the remaining four degrees of freedom.

In Fig. 4, we summarize the results of an extensive parametric study involving $h$ and $DC$, i.e., the parameters affecting the grating diffraction efficiency at a specific wavelength. For each pair, $d_t$ was determined in order to obtain a resonance around $\lambda_0$. The number of periods was fixed at $N = 5$. For the selected values of $h$ ($60\,\text{nm}$ and $30\,\text{nm}$), we observe shallower but much narrower resonances by comparison with the case in Fig. 3(b). In particular, by decreasing the grating height, the resonance lineshape tends to shrink (see the insets), but its visibility in the spectral range decreases as well. A similar visibility-bandwidth trade-off occurs also for the selection of the duty-cycle $DC$: for values approaching 0.5, the resonance becomes more visible in spite of a spectral broadening; conversely, far from 0.5, the spectral selectivity is enhanced, but is accompanied by an amplitude decrease.

We repeated the same study also for different values of $N$, observing the same behavior described above. More specifically, we observed that the number of periods does not sensibly affect the resonance linewidth (basically established by the values chosen for $d_t$, $h$ and $DC$), but it can lead to a significant increase of the resonance amplitude.

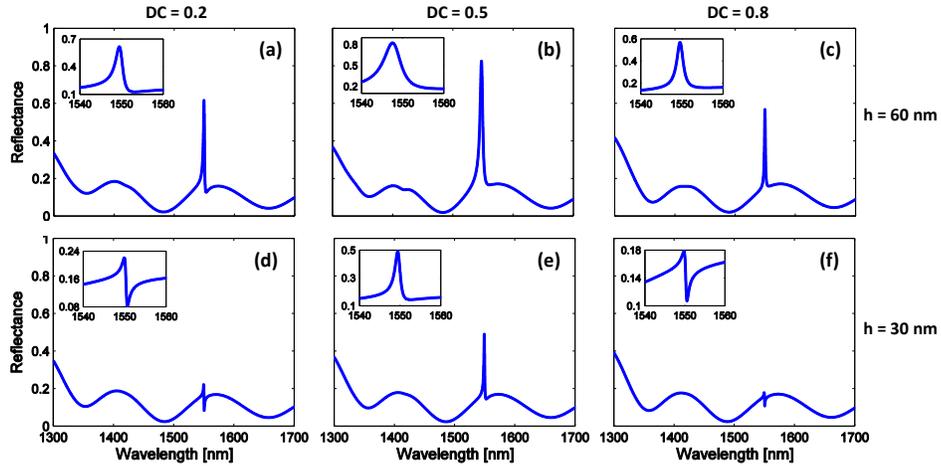

Fig. 4. Zeroth-order reflectance spectra for different values of $h$, $DC$ and $d_t$. The number of bilayers $N$ is fixed at $5$, and the remaining 1DPC parameters are as in Fig. 3. In the insets, the regions around the BSW resonances are magnified. (a) $h = 60\,\text{nm}$, $DC = 0.2$, $d_t = 72\,\text{nm}$; (b) $h = 60\,\text{nm}$, $DC = 0.5$, $d_t = 62\,\text{nm}$; (c) $h = 60\,\text{nm}$, $DC = 0.8$, $d_t = 54\,\text{nm}$; (d) $h = 30\,\text{nm}$, $DC = 0.2$, $d_t = 82\,\text{nm}$; (e) $h = 30\,\text{nm}$, $DC = 0.5$, $d_t = 77\,\text{nm}$; (f) $h = 30\,\text{nm}$, $DC = 0.8$, $d_t = 73\,\text{nm}$.

As a proof of this concept, we started considering the spectrum in Fig. 4(d) ($d_t = 82\,\text{nm}$, $h = 30\,\text{nm}$, $DC = 0.2$), which exhibits a shallow resonance with a desired narrow width ($\Gamma \approx 0.38\,\text{nm}$), and gradually increased $N$. As it can be observed in Fig. 5, the reflectance around the operational wavelength accordingly increases (leading to a consequent strong visibility enhancement), and is accompanied by negligible linewidth variations (see the insets). For a more quantitative assessment, we report in Table 1 the peak/dip excursion $H$ and the linewidth $\Gamma$ for each case.

To sum up, by dimensioning a poorly efficient diffraction grating, and gradually increasing the number of 1DPC periods (thereby reducing the radiation losses in the fiber associated to the Bragg modes of order $m = \pm 1$), a highly selective resonant condition can be eventually reached for which nearly 100% of the power couples back with the zeroth-order reflected mode, leading to optimum characteristics of the resonant lineshape. In our case, 7 SiO/SiO$_2$ bilayers guarantee a high-$Q$ ($\sim 5340$), strongly visible BSW resonance. In principle, even smaller values for $N$ would be needed if higher-contrast dielectric materials were used for the 1DPC, thereby enhancing its efficiency as a dielectric mirror.

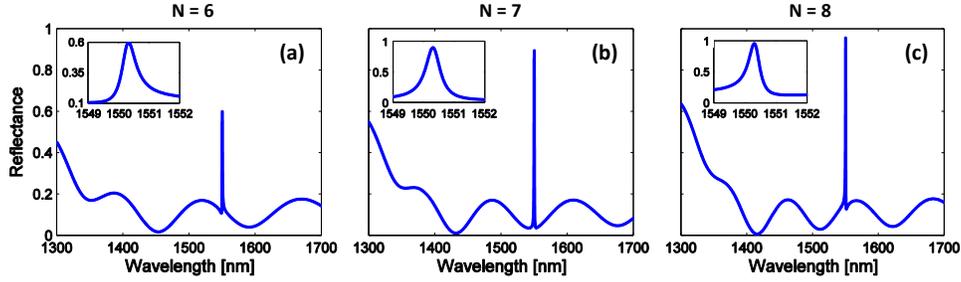

Fig. 5. Zeroth-order reflectance spectra for different values of $N$. All other parameters are as in Fig. 4(d). (a) $N = 6$; (b) $N = 7$; (c) $N = 8$.

Table 1. Numerical estimates of the peak/dip excursion ($H$) and the linewidth ($\Gamma$) of the BSW resonances obtained for four different values of $N$. All other parameters are as in Fig. 4(d).

| $N$ | $H$ | $\Gamma$ [nm] |
|---|---|---|
| 5 | 0.13 | 0.38 |
| 6 | 0.49 | 0.28 |
| 7 | 0.86 | 0.29 |
| 8 | 0.83 | 0.21 |

## 4. Sensing performance of the BSW tip

Since the main scope of this study is to identify the design rules for BSW-based label-free bioprobes integrated on optical-fiber tips, we move onto exploring how to design the final structure so as to tailor and enhance its performance as a refractive-index sensor. To this aim, as a meaningful observable, we consider the bulk sensitivity $S_B$, defined as the BSW resonance wavelength shift in the reflectance spectrum with respect to variations of $n_{ext}$. Since the smaller the difference between the coupling angle $\bar{\theta}_s$ and the critical angle $\theta_c$, the stronger the field exponential tail in the external medium (and, hence, its interaction with the environment), it is convenient to move the excitation angle $\bar{\theta}_s$ towards $\theta_c$. For simplicity, by still considering air ($n_{ext} = 1$) as the bulk region, we accordingly select $\bar{\theta}_s = 44°$.

By repeating the design procedure along the same line as the previous design in Secs. 3.1 and 3.2, we now obtain: $d_1 = 246$ nm, $d_2 = 368$ nm, $d_t = 0$ nm, $N = 8$, $\Lambda = 1522$ nm, $h = 35$ nm and $DC = 0.7$. The general applicability of the adopted method allows us to easily tailor the design in any other environments (e.g., water). Here, we basically provide a proof of principle of the sensing capabilities of the proposed device.

It is worth pointing out that, in the TE band diagram in Fig. 6 (computed for an infinite SiO/SiO$_2$ 1DPC with layer thicknesses $d_1$ and $d_2$), the position of the designed BSW (with coordinates $(\bar{k}_x, \bar{E})$, marked as black dot), corresponding to the surface mode supported at the interface between the semi-infinite 1DPC and air, is located around the center of the forbidden band. Here, $\bar{E} = 2\pi\hbar c/\lambda_0$ is the photon energy associated with the wavelength $\lambda_0$, with $\hbar$ denoting the Planck constant, and $c$ the speed of light in vacuum. This condition, attained also for the previous design (as a consequence of the 1DPC design procedure), guarantees the maximum BSW field attenuation in the 1DPC, and is desirable for sensitivity enhancement [18,19]. As it can be observed, the BSW is located at $\bar{k}_x = 1.018 k_0$, i.e., right below the air light-line.

In Fig. 7(a), we show the reflectance spectrum for the designed structure. A sufficiently sharp ($\Gamma \approx 0.82\,\text{nm}$, $Q \approx 1900$) and visible ($H \approx 0.6$) lineshape around $\lambda_0$ is attained. As before, a more visible resonance would arise by further increasing $N$. Indeed, we verified that the optimum case ($H \approx 0.93$) is achieved for $N = 11$. On the other hand, however, this would require the deposition of a much thicker structure on the fiber tip (each 1DPC unit cell has a thickness of $614\,\text{nm}$). Nonetheless, the performance estimation carried out hereafter is not affected by this particular choice, since the number of periods has a negligible influence on the bulk sensitivity [20,21].

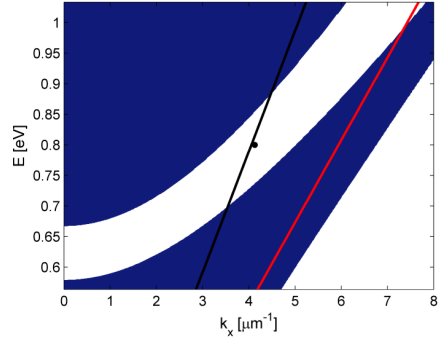

Fig. 6. Band diagram in the $(k_x, E)$ plane for an infinite 1DPC (TE polarization) with $d_1 = 246\,\text{nm}$ and $d_2 = 368\,\text{nm}$. The BSW location is indicated as a black dot, right below the air light-line (black curve) and above the silica light-line (red curve).

The field-map in Fig. 7(b) highlights the expected increase of the evanescent field content in the external medium at the operational wavelength with respect to one of Fig. 7(c) (related to previous design). In particular, by evaluating in both cases the surface integral of the electric field magnitude ($\iint |E(x,z)|\,dx\,dz$) inside the air domain (including the grating grooves and the external region) within a unit cell, we estimated an enhancement factor of $\sim 35\%$.

*4.1 Performance evaluation*

Figure 8 shows the sensitivity curve $\Delta\lambda_{BSW}(n_{ext})$ pertaining to the structure in Figs. 7(a) and 7(b), obtained by tracking the wavelength corresponding to the maximum slope of the BSW resonance for increasing values of the external medium refractive index $n_{ext}$, around its initial value $n_{ext} = 1$. An essentially increasing linear trend is observed for $n_{ext}$ variations up to $0.4 - 0.5\%$ (see the inset), with bulk sensitivity $S_B \approx 560\,\text{nm/RIU}$. For larger $n_{ext}$ variations, $\Delta\lambda_{BSW}$ starts increasing more than linearly until, for $\Delta n_{ext} > 1\%$, the nonlinearity becomes fairly evident. For $\Delta n_{ext} = 4\%$, we obtain $S_B \approx 810\,\text{nm/RIU}$. This sensitivity increase can be explained by considering that, by increasing $n_{ext}$, the critical angle $\theta_c$ increases and, as we numerically verified, tends to approach the coupling angle $\theta_s(\lambda_{BSW})$, with a consequent further enhancement of the evanescent field interaction.

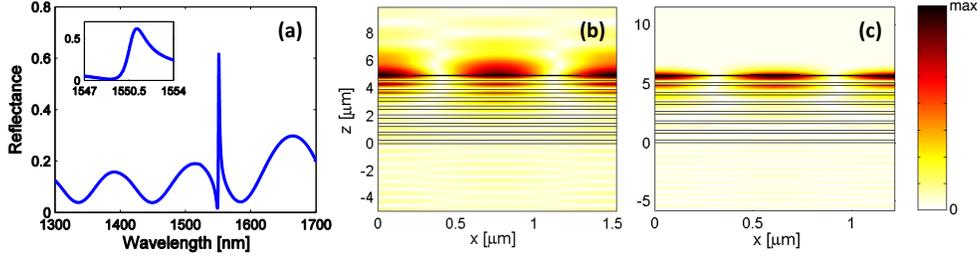

Fig. 7. (a) Zeroth-order reflectance spectrum for a structure as in Fig. 3(a) with $d_1 = 246\,\text{nm}$, $d_2 = 368\,\text{nm}$, $d_t = 0\,\text{nm}$, $N = 8$, $\Lambda = 1522\,\text{nm}$, $h = 35\,\text{nm}$ and $DC = 0.7$. In the inset, the region around the BSW resonance is magnified. (b) Electric field magnitude map within a unit cell, computed at the operational wavelength $\lambda_0 = 1550\,\text{nm}$. (c) As in panel (b), but for the structure pertaining to Fig. 5(b).

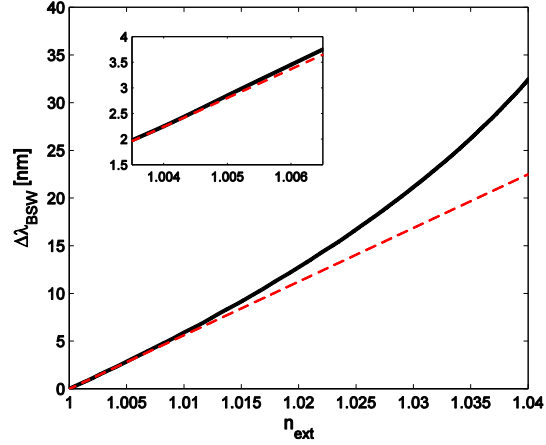

Fig. 8. Sensitivity curve $\Delta\lambda_{BSW}(n_{ext})$ (black-solid) pertaining to the structure in Figs. 7(a) and 7(b). The red-dashed line has a slope corresponding to the linear-regime bulk sensitivity $S_B \approx 560\,\text{nm/RIU}$. In the inset, the region where the non-linearity starts is magnified.

By repeating the same computation for the previous configuration (in which the BSW was coupled far from the cut-off condition), we evaluated in the linear regime $S_B \approx 130\,\text{nm/RIU}$, i.e., a sensitivity smaller by over a factor four.

Use of higher-contrast dielectric materials for the 1DPC could in principle lead to larger sensitivities. However, the calculated values are of the same order of magnitude of those reported for the BSW sensors proposed in [14,16,19].

### 4.2 Comparisons with lab-on-fiber plasmonic bioprobes

Overall, the proposed configuration shows intriguing potentials as a high-performance optical-fiber tip label-free biosensor, as indicated by a direct comparison with state-of-the-art plasmonic lab-on-tip probes.

Indeed, our numerically-estimated refractive-index sensitivities are of the order of hundreds of nm/RIU, as those exhibited by most structures proposed in the literature, mainly based on localized SPRs [29,34,35,47–49] (see Table 2). In [50], metallo-dielectric crystals fabricated via self-assembly were found to yield localized SPRs with surprisingly high sensitivities (up to $\sim 2300\,\text{nm/RIU}$) accompanied, however, by $Q$-factors lower than 15. Conversely, a very narrow linewidth of 6.6 nm ($Q \approx 130$) was attained in [51] for a SPR excited in a gold hexagonal nanohole array, with sensitivity once again in line with those reported above. Against this background, our results indicate the possibility to obtain a dramatic $Q$-factor enhancement, while still relying on a highly sensitive fiber tip detection scheme.

Table 2. Benchmark values of bulk sensitivity ($S_B$) and $Q$-factor reported in literature for state-of-the-art fiber tip plasmonic bioprobes.

| Ref. | Fiber tip structure | $S_B$ [nm/RIU] | Q |
|---|---|---|---|
| [29] | 2D hybrid metallo-dielectric nanostructure | 424.4* | 38* |
| [34] | | 125 | 23 |
| [35] | Array of gold nanodots | 195.72 | - |
| [47] | Array of gold nanodisks | 226 | - |
| [48] | Metallic nanoparticles | 387 | - |
| [49] | Array of subwavelength apertures | 533 | - |
| [50] | Self-assembled metallo-dielectric crystal | 2300 | < 15 |
| [51] | Gold hexagonal nanohole array | 595 | ~ 130 |

*Numerically estimated values

## 5. Effect of dielectric losses

It is worth highlighting that lossless materials (i.e., zero values for the extinction coefficients $\kappa_1$ and $\kappa_2$) were assumed in the numerical simulations carried out. This represents an important difference with respect to prism-coupled configurations, for which lossy materials are instead necessary to observe a dip in the reflectance spectrum [19]. More specifically, the studies in [20,52] showed that the extinction coefficients values play a key role in determining the optimum compromise between depth and width of the BSW resonance. On the other hand, in our case, losses in the 1DPC essentially leads to a degradation of the BSW lineshape obtained in the lossless case. In [41], this kind of detrimental effect was numerically observed by introducing losses in the bio-solution at the interface with a lossless 1DPC.

The perturbation effect induced by non-zero values for $\kappa_1$ and $\kappa_2$ on the resonances obtained for the two analyzed configurations [spectra in Figs. 5(b) and 7(a)] is illustrated in Figs. 9(a) and 9(b), and quantified in Table. 3. In the former case [Fig. 9(a)], a strong visibility decrease ($\sim 50\%$) is already noticeable for $\kappa_{1,2} = 10^{-4}$ (red dotted line) and, by further increasing the coefficients, the resonance rapidly tends to disappear. Conversely, the linewidth increases slowly for increasing values of $\kappa_{1,2}$ (see Table 3).

The latter configuration [Fig. 9(b)], given the stronger field enhancement in the external medium, turns out to be more robust against material losses in the 1DPC. As a matter of fact, the original BSW resonance is only slightly perturbed for $\kappa_{1,2} = 10^{-4}$ (red dotted line) and remains visible and fairly narrow ($\Gamma \approx 1.4\,\mathrm{nm}$) even for $\kappa_{1,2} = 10^{-3}$ (green dashed line).

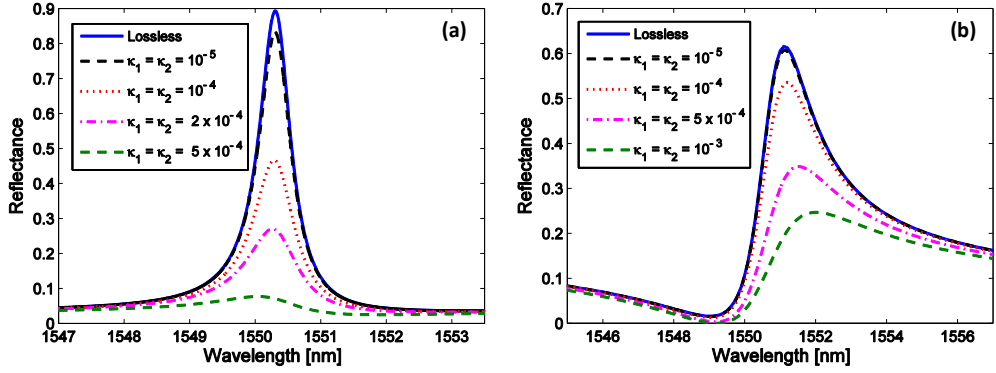

Fig. 9. (a) Variations of the BSW resonance lineshape for the configuration in Fig. 5(b), and different values of the extinction coefficients $\kappa_1$ and $\kappa_2$. (b) Same as panel (a), but for the configuration pertaining to Fig. 7(a).

Table 3. Numerical estimations of the peak/dip excursion ($H$) and the linewidth ($\Gamma$) of BSW resonances obtained for different values of the extinction coefficients $\kappa_{1,2}$ and for the configurations in Fig. 5(b) and 7(b).

|  | First design | | Second design | |
| --- | --- | --- | --- | --- |
| $\kappa_1 = \kappa_2$ | $H$ | $\Gamma$ [nm] | $H$ | $\Gamma$ [nm] |
| 0 | 0.86 | 0.29 | 0.60 | 0.82 |
| $10^{-5}$ | 0.80 | 0.30 | 0.59 | 0.82 |
| $10^{-4}$ | 0.43 | 0.38 | 0.52 | 0.88 |
| $2\times10^{-4}$ | 0.24 | 0.46 | 0.46 | 0.93 |
| $5\times10^{-4}$ | 0.05 | 0.70 | 0.34 | 1.10 |
| $10^{-3}$ | 0.03 | 1.06 | 0.25 | 1.40 |

## 6. Conclusions

To sum up, we have demonstrated the feasibility of a BSW-based label-free optical optrode that can be integrated on the tip of an optical fiber. More specifically, we have first explored the possibility to efficiently excite BSWs at the surface of a truncated 1DPC deposited on the fiber tip, overcoming the normal-incidence-induced limitations, as well as expectable fabrication non-idealities. This has led us to explore a coupling mechanism based on a 1D diffraction grating realized on the top of the 1DPC. Moreover, we have identified the degrees of freedom available in the design, and have exploited them in order to tailor and optimize the parameters of the arising resonance (visibility, linewidth) for specific applications. Subsequently, we have presented a design oriented to refractive-index sensing applications, by enhancing the evanescent field interaction with the external environment and then the bulk sensitivity. The sensitivity curve indicates a bulk sensitivity of $\sim 560\,\mathrm{nm/RIU}$ in the linear regime ($0.4 - 0.5\%$ $n_{ext}$ variations). Interestingly, in this case, a remarkable robustness of the arising resonance lineshape against material losses in the 1DPC was verified, even for extinction coefficients $\sim 10^{-3}$.

Clearly, an experimental verification of our numerical predictions is crucial to ultimately assess the capability of the proposed platform to outperform state-of-the-art plasmonic lab-on-fiber probes. In particular, a broadening of the BSW resonance is expectable, due to the finite size of the structure and the numerical aperture of the fiber, as well as the effects of fabrication tolerances. Accordingly, the fabrication and experimental characterization of prototypes are under way, and will be the subject of a future study.